\pdfoutput=1 
\documentclass[preprint,10pt,nocopyrightspace]{sigplanconf}
\usepackage{tikz}
\usetikzlibrary{trees,snakes}
\usepackage[ruled,vlined,linesnumbered]{algorithm2e}
\SetAlFnt{\small}
\SetAlCapFnt{\small}
\SetAlCapNameFnt{\small}

\usepackage{mathtools}

\usepackage{amsmath}

\usepackage{tikz}
\usepackage{ifthen}

\usepackage{listings}

\definecolor{turquoise}{rgb}{0.0,0.4,0.4}
\definecolor{commentcolor}{rgb}{0.0,0.4,0.4}
\definecolor{grey}{rgb}{0.4,0.4,0.4}
\definecolor{magenta}{rgb}{0.58,0,0.82}
\definecolor{orange}{rgb}{0.88,0.3,0.0}
\definecolor{keywordcolor}{rgb}{0.88,0.3,0.0}
\newenvironment{Code}
  {\noindent\begin{minipage}{\linewidth}}{\end{minipage}}

\usepackage{epstopdf}
\graphicspath{{./graphics/}{./graphics/}}
\DeclareGraphicsExtensions{.pdf,.jpeg,.png,.eps}

\usepackage[position=top,labelformat=empty]{subfig}
\newdimen\HilbertLastX
\newdimen\HilbertLastY
\newcounter{HilbertOrder}

\def\DrawToNext#1#2{%
   \advance \HilbertLastX by #1
   \advance \HilbertLastY by #2
   \pgfpathlineto{\pgfqpoint{\HilbertLastX}{\HilbertLastY}}
}

\def\Hilbert[#1,#2,#3,#4,#5,#6,#7,#8] {
  \ifnum\value{HilbertOrder} > 0%
     \addtocounter{HilbertOrder}{-1}
     \Hilbert[#5,#6,#7,#8,#1,#2,#3,#4]
     \DrawToNext {#1} {#2}
     \Hilbert[#1,#2,#3,#4,#5,#6,#7,#8]
     \DrawToNext {#5} {#6}
     \Hilbert[#1,#2,#3,#4,#5,#6,#7,#8]
     \DrawToNext {#3} {#4}
     \Hilbert[#7,#8,#5,#6,#3,#4,#1,#2]
     \addtocounter{HilbertOrder}{1}
  \fi
}

\def\hilbert((#1,#2),#3){%
   \advance \HilbertLastX by #1
   \advance \HilbertLastY by #2
   \pgfpathmoveto{\pgfqpoint{\HilbertLastX}{\HilbertLastY}}
   \setcounter{HilbertOrder}{#3}
   \Hilbert[1mm,0mm,-1mm,0mm,0mm,1mm,0mm,-1mm]
   \pgfusepath{stroke}%
}

\begin{document}

\setlength{\pdfpageheight}{\paperheight}
\setlength{\pdfpagewidth}{\paperwidth}

\conferenceinfo{CONF 'yy}{Month d--d, 20yy, City, ST, Country}
\copyrightyear{20yy}
\copyrightdata{978-1-nnnn-nnnn-n/yy/mm}
\copyrightdoi{nnnnnnn.nnnnnnn}


\preprintfooter{PROHA'16, March 12, 2016, Barcelona, Spain}

\title{Effective use of the PGAS Paradigm: Driving Transformations and
Self-Adaptive Behavior in DASH-Applications}

\authorinfo{Kamran Idrees}
           {High Performance Computing Center Stuttgart (HLRS)}
           {idrees@hlrs.de}
\authorinfo{Tobias Fuchs}
           {Ludwig-Maximilians-Universit{\"a}t M{\"u}nchen (LMU)}
           {tobias.fuchs@nm.ifi.lmu.de}           
\authorinfo{Colin W. Glass}
           {High Performance Computing Center Stuttgart (HLRS)}
           {glass@hlrs.de}               
           
\maketitle

\begin{abstract}
DASH is a library of distributed data structures and algorithms
designed for running the applications on modern HPC architectures, composed of
hierarchical network interconnections and stratified memory. 
DASH implements a PGAS (partitioned global address space)
model in the form of C++ templates, built on top of DART -- a run-time
system with an abstracted tier above existing one-sided communication
libraries.

In order to facilitate the application development process for exploiting 
the hierarchical organization of HPC machines,
DART allows to reorder the placement of the computational units. 
%
In this paper we present an automatic, hierarchical units mapping
technique (using a similar approach to the Hilbert curve transformation) 
to reorder the placement of DART units on the Cray XC40 machine 
Hazel Hen at HLRS. To evaluate the performance of
new units mapping which takes into the account the topology of
allocated compute nodes, we perform latency benchmark for a 3D stencil
code. The technique of units mapping is generic and can be be adopted
in other DART communication substrates and on other hardware
platforms.

Furthermore, high--level features of DASH are presented, enabling more complex 
automatic transformations and optimizations in the future.
\end{abstract}



\section{Introduction}

Partitioned Global Address Space (PGAS) devises a method of parallel 
programming by introducing a unified global view of address 
space (like purely shared memory systems) and presiding over the 
distribution of data (similar to a distributed memory system),
in order to provide a programmer with ease of use and a locality-aware paradigm. 
To distribute the data across the system, PGAS implementations use one--sided communication substrates, 
which are hidden from the application developer.

Unified Parallel C (UPC) is an implementation of the PGAS model. UPC has a single shared address space, which is partitioned among 
UPC threads, such that a portion of shared address space resides in the 
memory local to a UPC thread. UPC provides mechanisms to distinguish 
between local and remote data accesses, thus allowing to capitalize on 
data locality. 
However, a programmer needs to perform 
custom coding, potentially even building advanced 
data distribution schemes, to exploit locality efficiently.

DASH is a C++ library that delivers distributed data structures 
and algorithms designed for modern HPC machines, which are well-suited 
for hierarchical network interconnections and memory stratum \cite{furlinger2014dash}. 
DASH aims at various domains of scientific applications, providing the programmer with 
advanced data structures and algorithms, consequently reducing the need for custom coding.

DASH calls DASH Run-Time (DART), which provides basic functionalities
to build a PGAS model using state of the art one-sided communication 
libraries \cite{zhou2014dart}. These functionalities include: 

\begin{itemize}
 \item Global memory management and optimization for accessing data that reside on shared memory system \cite{zhou2015leveraging}
 \item Creation, destruction and management of teams and groups
 \item Collective and non-collective communication routines 
 \item Synchronization primitives
\end{itemize}

This paper presents DASH as an alternative to traditional PGAS 
implementations like UPC. Short-comings of UPC and other traditional 
PGAS implementations are discussed, which in many cases prevent an 
effective use of the PGAS paradigm. Furthermore, features of DASH 
overcoming these short--comings are presented.
The main contributions of this paper are: 

\begin{enumerate}
 \item We present an advanced local copy feature in DASH
  \item  We evaluate the throughput of the local copy feature
 \item We present an automatic hierarchical units mapping mechanism for applications having a nearest neighbor communication pattern
 \item We evaluate the applicability of automatic hierarchical units 
 mapping mechanism on a 3D stencil communication kernel on Cray XC40 
 Hazel Hen machine at HLRS 

\end{enumerate}

\section{Experiences with UPC}
\label{sec:UPC}
From our experience with UPC for our in--house molecular dynamics code, 
we highlighted three major issues, resulting in severe performance degradation \cite{idreesperformance}. 
These issues -- and a further problem regarding hardware topology -- are:

\begin{enumerate}
\item Manual pointer optimization is necessary for fast access to local data (using local pointer)
\item Non-trivial data distribution schemes need to be implemented by hand 
\item Communication is performed at the same granularity as data access 
\item No mechanism available for co--locating strongly interacting units on the given hierarchical hardware topology 
\end{enumerate}

The first issue regards the failure of UPC compilers to automatically distinguish between shared and 
distributed memory data accesses, even though the complete data layout is available. 
This holds true for both static and dynamic allocation of data (as the block 
size of a distributed shared array needs to be a compile--time constant). This can 
lead to a significant performance degradation and can only be 
avoided by expert programmer intervention. Manual optimization requires checking all 
parts of the code where significant data accesses are performed and switching to local pointers
for local memory access. 

The second issue is that UPC provides only round robin and blocked data distribution schemes. 
These schemes are suboptimal for many applications featuring some sort of short range 
geometric data accessing patterns, e.g. stencil patterns. This will lead to a unnecessarily 
high amount of communication traffic and percentage of remote communication. 
To avoid this problem, the programmer has to write specific data mapping routines.

The third issue is associated with the communication granularity. 
In shared memory address space, the programmer can directly access 
and modify the shared data. The PGAS paradigm also provides these 
attributes for its global address space. However, accessing and 
modifying remote data is expensive, especially if it leads to many 
small communications. As UPC does not change the granularity, 
this often leads to a vast number of tiny communications. To avoid 
this problem, the programmer needs to take the underlying distributed 
memory architecture into account and perform the necessary optimizations 
for packing communications manually.
 
The fourth issue addresses the difficulty of adapting the behavior of an 
application to the machine topology. For example, reordering the placement 
of software units that may allow to reduce the communication cost by 
placing the interacting partners closer to each other on the physical 
hardware.

To summarize: in order to achieve a near optimal performance using traditional PGAS 
implementations, the programmer needs to take care of a variety of issues manually. 
This contradicts the driving idea behind PGAS: ease of programmability.
The good news is, DASH is tackling these issues by providing automatic 
optimization for faster local data accesses \cite{zhou2015leveraging}, 
advanced data distribution schemes \cite{DASH2015_MakePattern}, algorithm 
specific routines for pre-fetching and packing of data and automatic hierarchical units mapping. 
The following section provides a detailed illustration of these advance features.

\section{DASH as a Solution}

DASH resolves the short--comings of the traditional 
PGAS implementations with its automatic optimizations, advanced data 
structures and algorithms. We will now explain a few specific features 
of DASH which address the problems highlighted in the previous section.

\subsection{Fast Access to Local Data}
The automatic detection of local vs. remote data access in DASH demonstrates an 
effective use of PGAS paradigm: every data access is performed in the most efficient way available.
This automatic behavior is achieved by capitalizing on the shared memory window feature of MPI-3 \cite{hoefler2012leveraging} used in the MPI version of DART (DART--MPI). 
The shared memory window can be accessed directly by local MPI processes 
using load/store operations (zero--copy model), allowing the processes to circumvent the single-copy model of the MPI layer. 
DART-MPI maps both global and shared memory windows to the same shared memory 
region, thus allowing the DART units on shared memory to directly access the local memory 
region. The DART units that are not part of the shared memory window, perform RMA operations using a global window. Furthermore, the use of the zero copy 
model for intra-node communication in DART scales down the memory bandwidth problem.
We have demonstrated in \cite{zhou2015leveraging}, our optimization of mapping both 
shared and global memory windows to the same memory region on shared memory, enables faster 
intra-node communication. 
This allows DASH programmers to iterate over distributed data structures without 
worrying about slow local data accesses, unlike UPC where manual 
pointer optimizations are necessary to avoid less efficient local data accesses \cite{idreesperformance}.

\subsection{High-Level Data Distribution Schemes}
\label{subsec:patterns}

DASH features several data distribution schemes (\emph{patterns}) that provide
highly flexible configurations depending on data extents and logical unit
topology. New pattern types are continuously added to DASH.
This flexibility leads to a large number of data distributions that can be
used for a single use case.

The preferable pattern configurations depend on the specific use case.
Algorithms operating on global address space strictly depend on domain
decomposition as they expect data distributions that satisfy specific
properties.

Without methods that help to configure data distributions, programmers must
learn about the differences between all pattern implementations and their
restrictions. We therefore provide high-level functions to automatically optimize data
distribution for a given algorithm and vice-versa.
These mechanism are described in detail in \cite{DASH2015_MakePattern}.
In this we present a classification of data distribution schemes based on
well-defined properties in the three mapping stages of domain decomposition:
partitioning, mapping, and memory layout.
This classification system serves two purposes:

\begin{itemize}
\item Provides a vocabulary to formally describe general data distributions by
      their semantic properties (\emph{pattern traits}).
\item Specifies constraints on expected data distribution semantics.
\end{itemize}

As an example, the \emph{balanced partitioning} property describes that data
is partitioned into blocks of identical size.
An algorithm that is optimized for containers that are evenly distributed
among units can declare the \emph{balanced partitioning} and
\emph{balanced mapping} properties as constraints. A mapping is balanced if
the same number of blocks is mapped to every unit.

When applying the algorithm on a container, its distribution is then checked
against the algorithm's constraints already at compile time to preventing
inefficient usage:

\begin{Code}
\begin{lstlisting}[caption=Checking distribution constraints at compile time
                           and run time.,
                   label=lst:constraints-checking]
static_assert(
   dash::pattern_contraints<
       pattern,
       partitioning_properties<
         partitioning_tag::balanced >,
       mapping_properties<
         mapping_tag::balanced >,
       layout_properties< ... >
   >::satisfied::value);
\end{lstlisting}
\end{Code}

Finally, pattern traits also allow to implement high-level functions that
resolve data distribution automatically.
For this, we use simple constrained optimization based on type traits in
C++11 to create an instance of a initially unspecified pattern type that is
optimized for a set of property constraints.
To be more specific, the automatic resolution of a data distribution involves
two steps: at compile time, the pattern type is deduced from constraints that
are declared as type traits.
Then, an optimal instantiation of this pattern type is resolved from
distribution constraints and run-time parameters such as data extents and team
size.

\begin{Code}
\begin{lstlisting}[caption=Deduction of an Optimal Distribution using dash::make\_pattern.,
                   label=lst:make-pattern]
TeamSpec<2> logical_topo(16, 8);
SizeSpec<2> data_extents(extent_x, extent_y);
// Deduce pattern:
auto pattern =
  dash::make_pattern<
    partitioning_properties<
      partitioning_tag::balanced >,
    mapping_properties<
      mapping_tag::balanced >
  >(sizespec, teamspec);
\end{lstlisting}
\end{Code}

Deduction of data distribution can also interact with team specification to
find a suitable logical Cartesian arrangement of units. As for domain
decomposition, DASH provides traits to specify preferences for logical team
topology such as ``compact" or ``node-balanced".
As a result, application developers only need to state a use case, such as
DGEMM, and let unit arrangement and data distribution be resolved
automatically.


While no automation can possibly do away with the need for manual optimization
in general, automatic deduction as provided by DASH greatly simplifies finding
a configuration that is suitable as a starting point for performance tuning.
In comparison, finding practicable blocking factors and process grid extents
for ScaLAPACK, even in seemingly trivial use cases, is a challenging task for
non-experts.

\subsection{Creating Local Copies}
\label{subsec:local-copy}

Algorithms and container types in DASH follow the concepts and semantics of
their counterparts in the C++ Standard Template Library (STL) and are
consequently based on the iterator concept.
Algorithms provided by the STL can also be applied to DASH containers and most
have been ported to DASH providing identical semantics.
Programmers will therefore already be familiar with most of the API concepts.
For copying data ranges, the standard library provides the function interface
\texttt{std::copy}. In DASH, a corresponding interface \texttt{dash::copy} is
provided for copying data in PGAS.

This section presents the concept of the the functions \texttt{dash::copy}
and \texttt{dash::copy\_async}, first-class citizens in the DASH algorithm
collection which represent a uniform, general interface for copy operations
within global address space. 

As an example, compare how an array segment is copied using standard library
and using DASH:

\begin{Code}
\begin{lstlisting}
double range_copy[ncopy];
// STL variant:
std::copy(arr.begin(), arr.begin() + ncopy,
          range_copy);
// DASH variant, on a global array:
dash::copy(arr.begin(), arr.begin() + ncopy,
           range_copy);
\end{lstlisting}
\end{Code}

Asynchronous variants of data movement operations are essential to enable
overlap of communication and computation.
They employ the \emph{future} concept also known from the standard library:

\begin{Code}
\begin{lstlisting}
double range_copy[ncopy];
auto fut_copy_end = dash::copy_async(
                      arr.begin(),
                      arr.begin() + ncopy,
                      range_copy);
// Blocks until completion and returns result:
double * copy_end = fut_copy_end.wait();
\end{lstlisting}
\end{Code}

Copying data from global into to local memory space is a frequent operation in
PGAS applications. It can involve complex communications pattern as some
segments of the copied range might be placed in memory local to the requesting
unit's core while others are owned by units on distant processing nodes.


Performance of copy operations in partitioned global address space is optimized
by avoiding unnecessary data movement and scheduling communication such that
interconnect capacity is optimally exploited.
We use the following techniques in DASH, among others:

\begin{description}
\item[Shared memory] For segments of the copied data range that are located
     on the same processing node as the destination range, \texttt{std::copy}
     is used to copy data in shared memory. This reduces calls to the
     communication back-end to the unavoidable minimum. 
     This is not restricted to copying: DASH algorithms in general automatically
     distinguish between accesses in shared and distributed memory.
     And even when not using DASH algorithms, the DASH runtime automatically
     resorts to shared window queries and \texttt{memcpy} instead of MPI
     communication primitives for data movement within a processing node.
\item[Chunks] To achieve optimal throughput between processing nodes,
     communication of data ranges is optimized for transmission buffer sizes
     by splitting the data movement into chunks.
     Adequate chunk sizes are obtained from auto tuning or interconnect
     buffer sizes provided in environment variables by some MPI runtimes.
\item[Communication scheduling] Parallel transmission capacity is exploited
     whenever possible: if the data source range spans the address space of
     multiple units, separate asynchronous transmissions for every unit are
     initiated instead of a sequence of blocking transmissions.
     Also, the single asynchronous operations are then ordered in a schedule
     such that communication is balanced among all participating units to fully
     utilize interconnect capacity.
\end{description}

The shared memory optimization technique require means to logically partition
a global data range into local and remote segments.
For this, DASH provides the utility function \texttt{dash::local\_range} that
partitions a global iterator range into local and remote sub-ranges:

\begin{Code}
\begin{lstlisting}
dash::Array<T> a(SIZE);
// Get iterator ranges of local segments in
// global array:
std::vector<dash::Range<T>> l_ranges =
  dash::local_range(a.begin() + 42, a.end());
// Iterates local segments in array:
for (l_range : l_ranges) {
  // Iterates values in local segments:
  for (l_value : l_range) {
    // ...
  }
}
\end{lstlisting}
\end{Code}

The global iterator range to partitioned into locality segments may be
multidimensional.
In addition, DASH containers provide methods to access blocks mapped to units
directly so that programmers do not have to resolve partitions from domain
decomposition themselves. 
For example, sub-matrix blocks can be copied in the following way:

\begin{Code}
\begin{lstlisting}
dash::Matrix<2, double> matrix;
// First block in matrix assigned to
// active unit, i.e. in local memory space:
auto l_block = matrix.local.block(0);
// Specify matrix block by global block
// coordinates, i.e. in global memory space:
auto m_block = matrix.block({ 3, 5 });
// Copy local block to global block:
dash::copy(l_block.begin(),
           l_block.end(),
           m_block.begin());
\end{lstlisting}
\end{Code}

The \texttt{dash::copy} function interface and the \texttt{dash::Matrix}
concept greatly simplify the implementation of linear algebra operations.
Efficiency of the underlying communication is achieved without additional
effort of the programmer due to the optimization techniques presented in
this section.

\begin{figure}
\begin{center}
\includegraphics[]{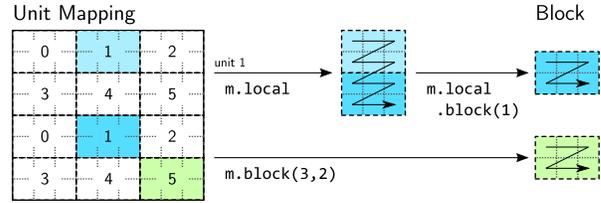}
\end{center}
\caption{Accessing blocks in a two-dimensional DASH matrix. The \texttt{local}
         access modifier changes the scope of the succeeding block access to
         local memory space.}
\label{fig:matrix-blocks}
\end{figure}

\subsection{Automatic Hierarchical Units Mapping}

The mapping of an application's software units (or threads/processes/tasks) 
to the physical cores on an HPC machine is becoming increasingly important 
due to the rapid increase in the number of cores on a machine \cite{hoefler2011generic}\cite{deveci2014exploiting}. 
This also 
leads to increasingly hierarchical networks and -- depending on the 
underlying system and the submitted job -- sparse core allocations. 
Therefore, if two units of an application which are interacting partners 
(or communicating more frequently than average) are placed far from each 
other in the network, they will have to communicate through several levels in the
network hierarchy. The placement of these units not only has repercussions in
their communication latency and bandwidth, but may also result in the
congestion of the network links.

\begin{figure}
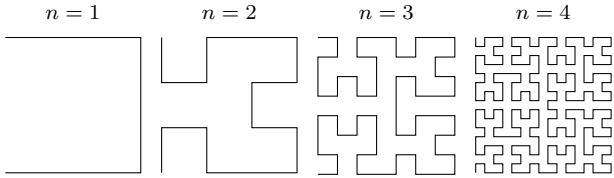
%
\begin{center}
    \subfloat[$n=1$]{\tikz[scale=18] \hilbert((0mm,0mm),1);}~~
    \subfloat[$n=2$]{\tikz[scale=6] \hilbert((0mm,0mm),2);}~~
    \subfloat[$n=3$]{\tikz[scale=2.6] \hilbert((0mm,0mm),3);}~~
    \subfloat[$n=4$]{\tikz[scale=1.2] \hilbert((0mm,0mm),4);}~~
\end{center}
\caption{Hilbert Space Filling Curve Example}
\label{hilbert}
\end{figure}

As the units mapping plays a vital role, DART-MPI provides DASH with the mean
to automatically reorder the units which respect both the communication 
pattern of an application and the topology of allocated nodes 
on an HPC machine. This results in reduced communication and 
overall execution time of an application. Currently we counter 
this problem for a specific set of applications which are based 
on nearest neighbor communication. The programmer only needs to 
inform DASH that the application to be executed has a nearest 
neighbor communication pattern. The automatic mapping routine 
then gathers the required hardware topology, computes a new 
hierarchical unit mapping and registers it in the system.
The new mapping is determined based on an approach similar to 
a Hilbert Space Filling Curve (HSFC) partitioning \cite{moon2001analysis}. 
An example of HSFC is shown in figure \ref{hilbert}. HSFC is chosen 
due to its property of preserving the locality. 
The algorithm however does not fix the length (number of elements to iterate 
over in a dimension) of HSFC for multiple levels
as the lengths are dependant upon the number of nodes 
corresponding to each level in the network hierarchy. 

Before discussing the steps in the automatic hierarchical unit mapping 
algorithm, we briefly explain the hierarchical network levels of the Cray XC40
Supercomputer Hazel Hen in the following.

\subsubsection{Off-Node Network Hierarchy on Hazel Hen}
\label{network-hierarchy}

The first off-node network hierarchical level is a \textit{compute blade}. A compute blade 
is composed of four nodes which share an Aries chip. The Aries 
chip connects these four nodes with the network interconnect. This is the fastest connection 
between nodes.

The second level is the \textit{rank 1 network} (or backplane). The rank 1 network is used for 
inter-compute blades communication within a chassis -- set of 16 compute blades (64 nodes). 
The rank 1 network has adaptive routing per packet. Any two nodes at this level communicate with 
each other by going first through their Aries chip, then through the backplane, and finally to
the Aries chip of the target node. This is an all-to-all PC board network.

The \textit{rank 2 network} is used for 
inter-backplane communication (nodes on distinct chassis) in a two cabinet group 
(a group is composed of 384 nodes). The backplanes are connected through copper cables. 
All copper and backplane signals run at 14 Gbps. The minimal route between two nodes 
on distinct chassis is two hops, whereas the longest route requires four hops. 
The aries adaptive routing algorithm is used to select the best route from four routes 
in a routing table. The rank 2 network also has an all-to-all connection, connecting 6 Chassis 
(two cabinets).

The last off-node network level is the \textit{rank 3 network}, which is used for 
communication between different groups. The rank 3 network has all-to-all routing 
using optical cables. If minimal path between two groups is congested, traffic can be 
hopped through any other intermediate group (1 or 2 hops).

\begin{figure}
\begin{center}
\includegraphics[scale=0.4]{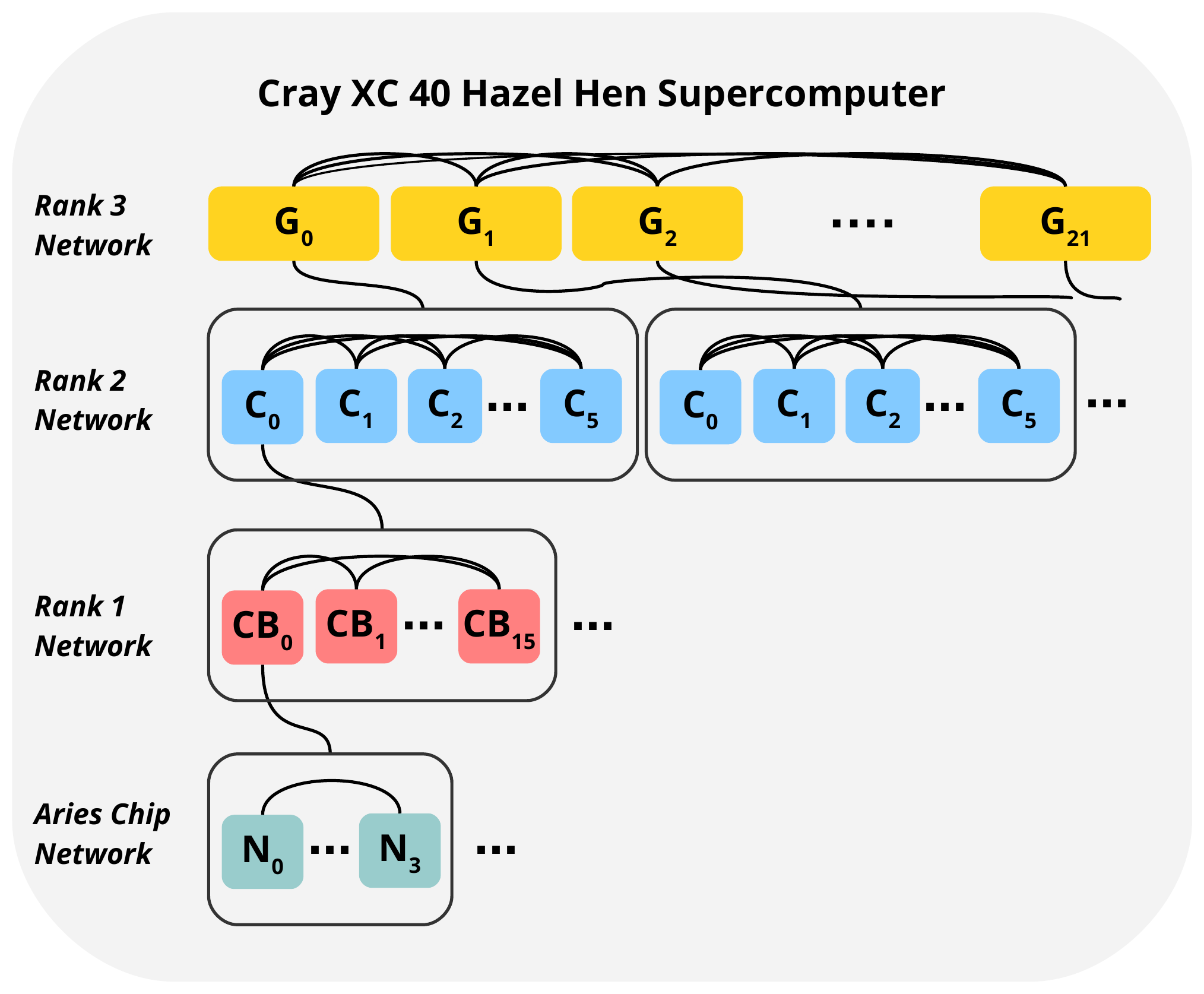}
\end{center}
\caption{Layered Layout of the Network Hierarchy of \texttt{Hazel Hen}. G, C, CB and N are 
abbreviated for Group, Chassis, Compute Blade and Node respectively.}
\label{hazel-hen}
\end{figure}

The layered layout of network hierarchy of Hazel Hen is show in figure \ref{hazel-hen}. 

\subsubsection{Algorithm}

The automatic hierarchical units mapping algorithm is executed by every DASH unit 
and assumes that the user has performed a binding of the units to the CPU cores, such that 
the units do not migrate from one CPU to another \footnote{On a Cray machine, unit 
or process binding can be easily performed by adding the argument \textit{-cc cpu} to the
\textit{aprun} command.}. Every unit performs the following steps:

\begin{enumerate}

\item Acquires the \textit{total number of units} in the team 
(to be mapped on the hierarchical topology).

\item Acquires its processor name and parses it to obtain the \textit{Node ID} 
on which the unit resides.

\item Participates in the collective allgather operation to obtain the Node IDs of all 
units (units on the same node have the same Node ID) and uses the Node ID  as a key to look for  
the placement information string of the node inside a topology file 
of the machine \footnote {A topology file on a Cray machine can be created using Cray's \textit{xtprocadmin} utility.
The placement information string of a node looks like \textit{c11-2c0s15n3}, which means 
the position of the node in the machine hierarchy is: column 11, row 2, chassis 0, 
compute blade 15 and node 3.}.

\item Reads topology file to acquire placement information string, 
number of sockets and number of cores per socket, for each allocated node.

\item Parses the placement information string of each node in order to obtain 
the value of each hierarchical level 
of the machine corresponding to each node. 

\item Sorts the nodes with respect to all levels in the \textit{network hierarchy} i.e. 
at first performing the sorting according to the values of every node on Level[4], 
then on
Level[3] and so on. For example:

Level(4, 3, 2, 1, 0) = (0, 0, 1, 12, 3)\newline
Level(4, 3, 2, 1, 0) = (0, 0, 1, 13, 0)\newline
Level(4, 3, 2, 1, 0) = (0, 0, 1, 13, 1)\newline
\vdots{}\newline
Level(4, 3, 2, 1, 0) = (9, 1, 2, 1, 1)

\item Determines balanced distribution of total number of units in a cartesian grid.

\item Performs balanced distribution of units per node, to form multi--core groups in 
order to reduce inter-node communication (For example: a balanced distribution for 24 cores as in Hazel Hen would be $4 \times 3 \times 2$. The lengths of coordinate directions 
of the 3D Cartesian grid (x,y,z) of total number of units should be divisible by the 
cartesian grid of units per node (e.g. (4,3,2)). This is necessary as 
our reordering method (Algorithm \ref{mapping-algorithm}) performs multi--core grouping at the node 
level and therefore the number of groups in each coordinate direction should fit the cartesian 
grid of total number of units.

\item Assigns new unit ID to each unit taking into consideration the multi--level network hierarchy, 
i.e. multicore groups of units are mapped as close as possible in the network hierarchy 
in order to reduce communication between distinct network hierarchy levels.

\item Finally, the reordered unit IDs are registered in the system.

\end{enumerate}

\begin{figure}
\begin{center}
\includegraphics[scale=0.35]{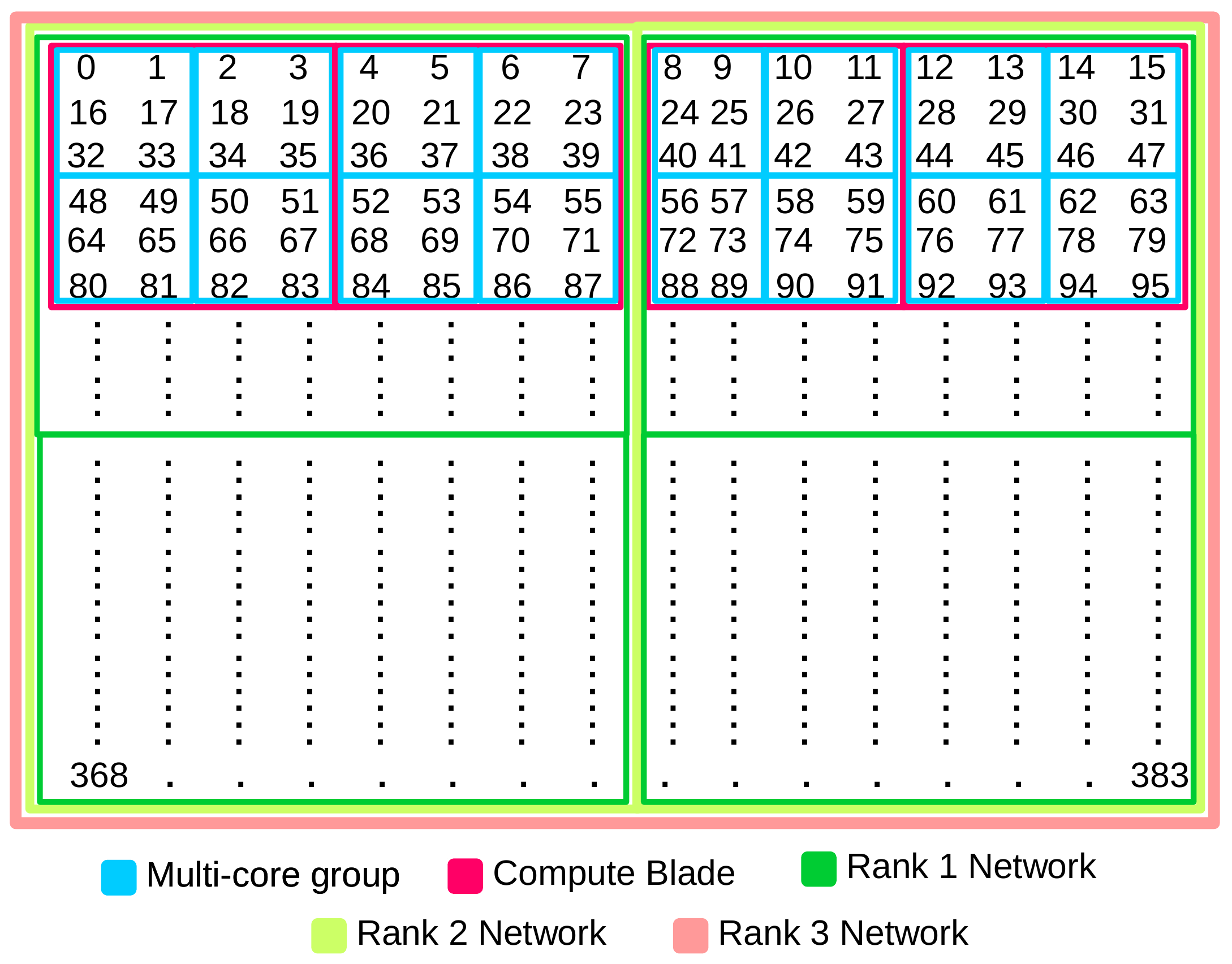}
\end{center}
\caption{Example of automatic hierarchical units mapping for a 2D nearest neighbor problem of size (24$\times$16). 
Please note that the example is just for elaborating the hierarchical units mapping and does not reflect the actual 
number of hardware instances at each network hierarchy level.}
\label{fig:units-mapping}
\end{figure}

After the last step, the new mapping is completed. The algorithm will result in an optimal units mapping if the 
node allocation is contiguous on all network hierarchy levels. Optimal being the minimal surface area, which results in minimal communication traffic on all network hierarchy 
levels. If the nodes are allocated in a sparse manner, the algorithm attempts to preserve 
the locality through its HSFC--like implementation.
Figure \ref{fig:units-mapping} shows an example of automatic hierarchical units mapping for 
a 2D nearest neighbor communication pattern.

%

\begin{algorithm}[]
\DontPrintSemicolon
\SetAlgoLined
\SetKwInOut{Input}{Input}
\SetKwInOut{Output}{Output}
\BlankLine
\Input{Balanced distribution of total number of units, number of units per node}
\BlankLine
\Output{Unit IDs respecting network hierarchy}
\BlankLine
\BlankLine 
$unitNumber \leftarrow 0$	\;
$N4 \leftarrow num Third Level Blocks In Fourth Level[3]$	\;
$N3 \leftarrow num Second Level Blocks In Third Level[3]$	\;
$N2 \leftarrow num First Level Blocks In Second Level[3]$	\;
$N1 \leftarrow num Nodes At First Level[2]$	\;
$N0 \leftarrow num Units Per Node[3]$	\;
$NT \leftarrow num Total Units[3]$ \;
\BlankLine
\tcc{Given that the nodes are sorted according to multiple network hierarchy levels, compute the new unit IDs}
\ForEach{x, y, z: 0 $\to$ N4 (0, 1, 2)}{	
\ForEach{a, b, c: 0 $\to$ N3 (0, 1, 2)}{	
\ForEach{d, e, f: 0 $\to$ N2 (0, 1, 2)}{	
\ForEach{g, h: 0 $\to$ N1 (0, 1)}{	
	\ForEach{i, j, k: 0 $\to$ N0 (0, 1, 2)}{
			\BlankLine
		\If{unitNumber $>$ totalUnits}{
			break \\
		}
		\BlankLine
		fourthLevelOffset $\leftarrow$  ..
		\BlankLine
		thirdLevelOffset $\leftarrow$  ..
		\BlankLine
		secondLevelOffset $\leftarrow$  d $\times$ N0[0] $\times$ NT[1] $\times$ NT[2] + e $\times$ N1[0] $\times$ N0[1] $\times$ NT[2] + f $\times$ N1[1]  $\times$ N0[2];
		\BlankLine
		firstLevelOffset $\leftarrow$ g $\times$ N0[1] $\times$ NT[2] + h $\times$ N0[2];
		\BlankLine
		\BlankLine
		unitID[unitNumber] $\leftarrow$ fourthLeveOffset + thirdLevelOffset + secondLevelOffset + firstLevelOffset + i $\times$ NT[1] $\times$ NT[2] + j $\times$ NT[2] + k;
		\BlankLine	
		unitNumber$++$;			
	}	
}
}
}
}
\tcc{Register the new unit IDs in run time system -- reordering}
\caption{Pseudo--code of algorithm to compute new unit IDs which respect underlying machine's network hierarchy}
\label{mapping-algorithm}
\end{algorithm}

\section{Performance Evaluation of Local Copy}

Optimization techniques employed in copying data in global memory space have
been discussed in \ref{subsec:local-copy}. 
In the following, we evaluate the \emph{local copy} use case where a unit
creates a local copy of a data range in global memory.
We consider the following scenarios, named by the location of the copied
data range:

\begin{description}
\item[local] Both source- and destination range are located at the same unit.
     This scenario does not involve communication as \texttt{dash::copy}
     resorts to copying data directly in shared memory.
     To illustrate the maximum achievable throughput, this scenario is also
     evaluated using \texttt{std::copy}.
\item[socket] The data range to be copied and the copy target range are owned
     by units mapped to different sockets on the same processing node. In this
     case, communication is avoided by DART recognizing the
     data movement as an operation on shared memory windows.
\item[remote] The source range is located on a remote processing unit and is
     copied in chunks using \texttt{MPI\_Get}.
\end{description}

For meaningful measurements, it is essential to avoid a pitfall regarding
cache effects: if the copied data range has been initialized by a unit placed
on the same processing node as the unit creating the local copy, the data
to be copied is stored in L3 data cache shared by both units.
In this case, the \texttt{local} and \texttt{socket} scenarios would
effectively measure cache bandwidth instead of the more common and less
convenient case where copied data is not available from cache.

\subsection{Benchmark Environment}

The local copy benchmark has been executed on SuperMUC phase 2 nodes for the
available MPI implementations Intel MPI, IBM MPI, and OpenMPI.
The MPI variants each exhibit specific advantages and disadvantages:

The installation of IBM MPI does not support MPI shared windows, effectively
disabling the optimization in the DASH runtime for the \emph{socket} scenario,
but offers the most efficient non-blocking RDMA.

Intel MPI requires additional polling processes for asynchronous RDMA which
increases overall communication latency.

The benchmark application has been compiled using the Intel Compiler (icc)
version 15.0.
Apart from being linked with different MPI libraries, the build environment
is identical for every scenario.

\subsection{Results}

\begin{figure*}%
\centering
\includegraphics[]{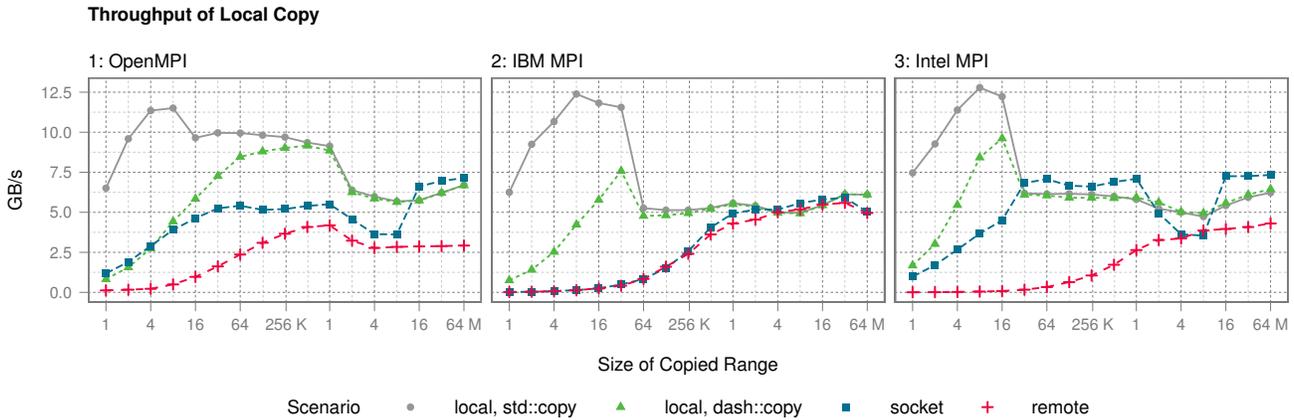}
\caption{Throughput of \texttt{dash::copy} for local copies of data ranges
         from different locality domains. Throughput of \texttt{std::copy}
         is included for reference.}
\label{fig:local-copy}
\end{figure*}

The results from all scenarios for the three MPI implementations is shown in
Figure \ref{fig:local-copy}.

As a first observation, performance of \texttt{std::copy} varies with the MPI
implementation used. This is because different C standard libraries must be
linked for the respective MPI library.
This also explains why cache effects become apparent for different range
sizes in the \emph{local} scenarios. In general, performance of local copying
is expected to decrease for ranges greater than 32 KB which is the capacity
of L1 data cache on SuperMUC Haswell nodes.
The C standard library linked for OpenMPI sustains better performance for
larger data sizes compared to the other evaluated MPI variants.

When copying very small ranges in the \emph{local} scenario the constant
overhead in \texttt{dash::copy} introduced by index calculations outweighs
communication cost. Still, the employed shared memory optimization leads to
improved throughput compared to MPI operations used in the \emph{remote}
scenario.

For copied data sizes of roughly 64 KB and greater, \texttt{dash::copy} achieves
the maximum throughput measured using \texttt{std::copy}.
This corresponds approximately to a minimum of a $90 \times 90$ block of double-precision
floating point values and is far below common data extents in
real-world use cases. 

As expected, achieved throughput in the \emph{socket} and \emph{remote}
scenarios are comparable for IBM MPI as shared window optimizations are not
available and MPI communication is used to move data within the same node.
Fortunately, IBM MPI also exhibits the best performance for MPI Communication.
For ranges of 1 MB and larger, there is no significant difference between
local and remote copying.

It might seem surprising that throughput in the \emph{socket} scenario, where
data is copied between NUMA domains, exceeds throughput in scenario
\emph{local} in some cases.  
However, in the \emph{socket} scenario, data is copied in the DASH runtime
using \texttt{memcpy} instead of \texttt{std::copy}.
The different low-level variants are expected to yield different performance
and again depend on the C standard library linked.

\begin{figure}%
\centering
\includegraphics[]{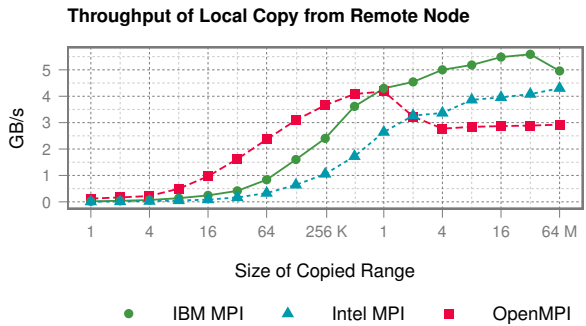}
\caption{Throughput of \texttt{dash::copy} for local copies of blocks on a remote processing
         node.}
\label{fig:local-copy-remote}
\end{figure}

Figure \ref{fig:local-copy-remote} summarizes achieved throughput in the
\emph{remote} scenario of all MPI implementations in a single plot for
comparison. 

Results from this micro-benchmark can serve to auto-tune partition sizes
used to distribute container elements among units. For example, a minimum
block size of 1 MB is preferable for IBM MPI while block sizes between 1 to
16 MB should be avoided for Intel MPI and OpenMPI as NUMA effects decrease performance
otherwise.

\section{Performance Evaluation of the Stencil Kernel}
\label{sec:eval-stencil}
We now evaluate the performance of 3D stencil communication kernel 
with and without using the automatic hierarchical units mapping feature. 
In order to measure solely the impact of hierarchical units mapping on the 
performance, we have disabled the shared memory window feature of 
DART-MPI for the benchmark shown later.

\subsection{Evaluation Metric}
In this stencil communication kernel, each unit communicates with
six neighbors (left, right, upper, lower, front, and back). We 
use the blocking DART put operation for transferring the messages 
from one unit to another and the size of the messages is varied exponentially 
from 1 byte to 2 megabytes. 

We are interested in evaluating the \textit{relative performance improvement 
factor}, which is computed by taking the ratio of the average execution 
times (ten--thousand iterations) of stencil communication kernel using default 
(as performed by the job launcher on the Cray machine) against hierarchical units mapping. 

\subsection{Benchmark Environment}
The benchmarks are carried out on the Cray XC40 machine Hazel Hen at HLRS. 
Each node on Hazel Hen is based on Intel Xeon CPU E5-2680 v3 (30M Cache, 
2.50 GHz) processors and comprises 24 cores (12 cores per socket). 
Cray's Aries interconnect provides node-node connectivity with multiple
hierarchical network levels. We have two on node memory hierarchy levels, which are 
Uniform Memory Access (UMA) -- intra-socket communication -- and Non-uniform 
Memory Access (NUMA) -- inter-socket communication. It's easy to exploit 
the on node hierarchical levels. In this paper we are more interested in 
showing results by exploiting the off-node network hierarchical levels 
(Section \ref{network-hierarchy}). 

\begin{figure}
\begin{center}
\includegraphics[scale=0.7]{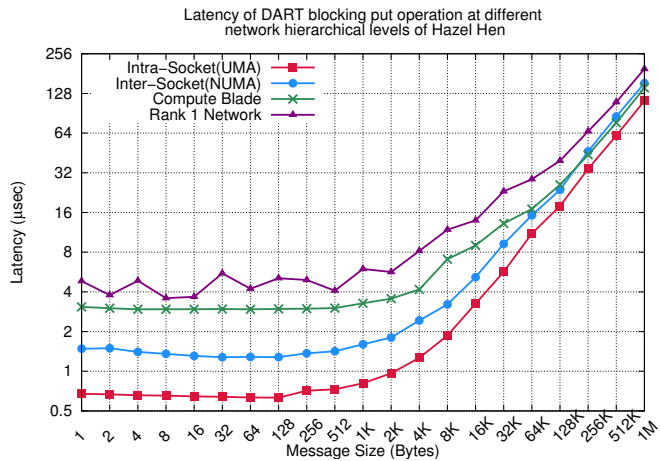}
\end{center}
\caption{Average latency of transferring a message between two nodes at different hierarchical levels of the network. 
The message size is varied from 1 Byte to 1 MegaByte.}
\label{hazel-hen-latency}
\end{figure}

\subsection{\textbf{Results}}
\label{sec:results}
Figure 
\ref{hazel-hen-latency} shows the average 
latency of messages up to the rank 1 network of Hazel Hen,  highlighting the impact of different network levels.

\begin{figure}
\begin{center}
\includegraphics[scale=0.7]{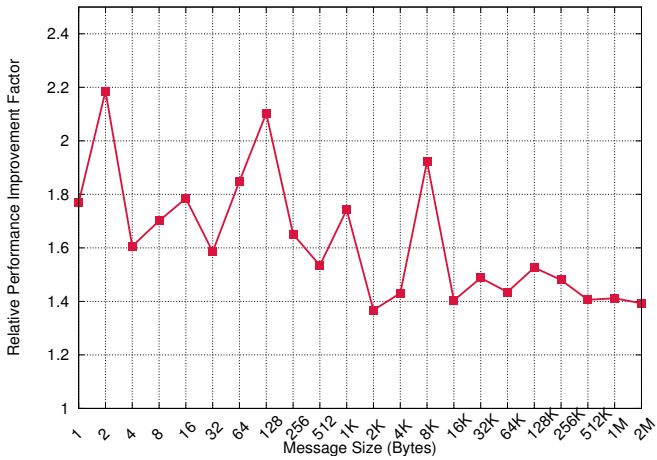}
\end{center}
\caption{Relative performance improvement factor of 3D stencil communication kernel using 
default against automatic hierarchical units mapping, on 384 sparse nodes on Cray Hazel Hen
XC40 machine. The message size is varied from 1 Byte to 2 MegaBytes.}
\label{mapping-result}
\end{figure}

Figure \ref{mapping-result} shows the relative performance improvement factor of the
3D stencil communication kernel on 384 nodes (9,216 units) of Hazel Hen. The nodes 
were allocated in a sparse manner by Cray's job launcher, having small contiguous 
blocks of nodes. It can be seen that our units mapping provides an average performance 
improvement by a factor of 1.4 to 2.2.

\section{Conclusions}
In this paper we have presented specific features of DASH which resolve some issues we 
observed in traditional PGAS implementations. We have shown in section \ref{sec:eval-stencil} that 
our automatic hierarchical units mapping provides a notable performance improvement 
over default units mapping. A user can take advantage of this self--adapting behavior 
without putting any effort into understanding the complex machine hierarchy or performing any custom 
coding. 
Furthermore, new features are presented enabling a user 
to represent the computation and communication patterns of scientific applications at 
a very high level of abstraction in DASH, while DASH will take care of necessary code transformations.
We are currently working on extending methods for automatic data distribution to data flow 
scenarios. However, automatic optimization in many data flow use cases is conceptually
equivalent to integer programming and thus proven to be NP-hard.
We assume that solutions for a useful subset of scenarios can be found using linear
programming techniques like the Simplex algorithm.



\acks

This work was supported by the project \emph{DASH} which is funded by the German Research Foundation (DFG) 
under the priority program "Software for Exascale Computing - SPPEXA" (2013-2015).


\bibliographystyle{abbrvnat}

\end{document}